\documentclass[preprint]{article}
\usepackage{eurosym}
\usepackage{amsmath,amssymb,amsthm}
\usepackage{graphicx}
\usepackage{subcaption}
\usepackage{verbatim}
\usepackage{bm}
\usepackage{color}
\usepackage{amsmath}
\usepackage{amsfonts}
\usepackage{amssymb}
\usepackage[margin=1in]{geometry}

\begin{document}

\title{Not the garden hose instability: wavelength selection in a buckling
garden hose}
\author{Tianyi Guo$^{1}$ \thanks{
Email: tguo2@kent.edu}, Xiaoyu Zheng$^{2}$\thanks{
Email: xzheng3@kent.edu}, Peter Palffy-Muhoray$^{1,2}$\thanks{
Corresponding author. Email: mpalffy@kent.edu} \\
\emph{$^1$Advanced Materials and Liquid Crystal Institute, Kent State
University, OH, USA}\\
\emph{$^2$Department of Mathematical Sciences, Kent State University, OH, USA%
} }
\maketitle

\begin{abstract}
We consider sinusoidal undulations which appear on certain garden hoses
under normal use. We propose a model, using linear elasticity, explaining
this phenomenon, and make a connection with biological structures as well as
self-buckling. We compare observations with model predictions, and suggest
potential applications in the area of shape-changing materials.
\end{abstract}

\section{Introduction}

\begin{figure}[th]
\centering
\includegraphics[width=.4\linewidth]{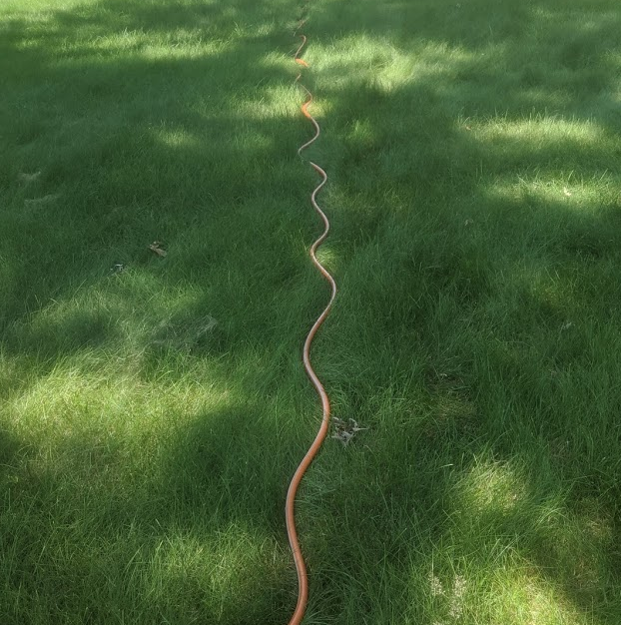}
\caption{ Example of undulations in a garden hose.}
\label{fig_1}
\end{figure}

An intriguing summertime phenomenon is the sinusoidal shape sometimes
assumed by certain garden hoses left on lawns. One example is shown in Fig.\ %
\ref{fig_1}. The pattern, which can extend over the entire length of the
hose, is apparently spontaneously formed during normal use. We have studied
this curious phenomenon and found unexpected connections with living
organisms. In this paper, we consider the underlying physics, and examine
the threshold for buckling, both in the garden hose and in a biological
system. We compare experimental results with model predictions for
wavelength selection in the case of our garden hose instability.

\section{The appearance of the deformation}

Garden hoses are typically connected to a pressurized domestic water system
via a shutoff valve at the inlet end, and are terminated with a spray nozzle
with a control valve at the outlet end. In the normal process of watering,
it is first ensured that the control valve by the nozzle is closed. Next,
the inlet valve is opened; at this point, water flows into the hose,
pressurizing it. The outlet valve is then opened; water flows through the
hose and sprays out of the nozzle, which is directed at whatever is to be
watered. When watering is completed, the outlet valve by the nozzle is
typically closed first, followed by closing the inlet valve. At this point,
the hose, still pressurized, is lying on the grass, often in a fairly
straight line. If the outlet valve is now opened to relieve the pressure in
the hose, many hoses exhibit a remarkable phenomenon: they visibly elongate
as water is expelled and form sinusoidal undulations along their length.
The same phenomenon occurs due to simple leakage of water from the hose. Two
questions immediately arise: why does the hose elongate, and what is the
mechanism of wavelength selection? We attempt to answer both questions below.

\section{Garden hose architecture}

A traditional material for garden hoses, still a popular option today, is
rubber. Rubber garden hoses, exhibit neither the elongation nor the
sinusoidal undulations described above. Elementary considerations show that
the longitudinal stress $\sigma _{l}$ in a thin-walled rubber tube of radius
$r$ and wall thickness $w$, where $r\gg w$, under excess pressure $P$ is%
\begin{equation}
\sigma _{l}=P\frac{r}{2w},  \label{1}
\end{equation}%
while the azimuthal stress $\sigma _{a}$ is
\begin{equation}
\sigma _{a}=P\frac{r}{w}.  \label{2}
\end{equation}%
Remarkably, the azimuthal stress is twice the longitudinal one. \ The
average radial stress is $P/2$, which is negligible in thin-walled tubes.

To calculate the corresponding strains, we turn to Hooke's Law in compliance
form. If there are no shear stresses and the radial stress is negligible, we
have \cite{LL}%
\begin{equation}
\left[
\begin{array}{c}
\varepsilon _{l} \\
\varepsilon _{a}%
\end{array}%
\right] =\frac{1}{E}\left[
\begin{array}{cc}
1 & -\nu \\
-\nu & 1%
\end{array}%
\right] \left[
\begin{array}{c}
\sigma _{l} \\
\sigma _{a}%
\end{array}%
\right] ,
\end{equation}%
where $E$ is Young's modulus and $\nu $ is Poisson's ratio. Substitution
gives the longitudinal strain
\begin{equation}
\varepsilon _{l}=\frac{1}{E}\left( \frac{1}{2}-\nu \right) P\frac{r}{w}
\end{equation}%
and the azimuthal strain%
\begin{equation}
\varepsilon _{a}=\frac{1}{E}\left( 1-\frac{1}{2}\nu \right) P\frac{r}{w}.
\end{equation}%
For incompressible materials, $\nu =1/2$. Since rubber is nearly
incompressible, we find that $\varepsilon _{l}=0$, so that remarkably, there
is no appreciable longitudinal extension of thin-walled tubes of
incompressible materials when pressurized. Such behavior is desirable in
blood vessels; this point is discussed in some detail for both rubber tubes
and arteries in \cite{Gordon} and by us in Sec.\ 6.

A simple rubber hose thus expands or contracts in diameter when pressurized or when the pressure is removed, but remains essentially constant in
length. It is then interesting to ask why certain garden hoses show a
dramatically different behavior, and lengthen significantly when their
internal pressure is reduced.

With advances in materials technology, a wide variety of materials are
available for the construction of tubing in addition to rubber, such as
polyurethane, polypropylene, polyvinyl chloride and others. More to the
point, a variety of architectures are also available, and composite
structures with linings and reinforcing meshes are frequently employed.

\begin{figure}[h]
\centering
\includegraphics[width=.4\linewidth]{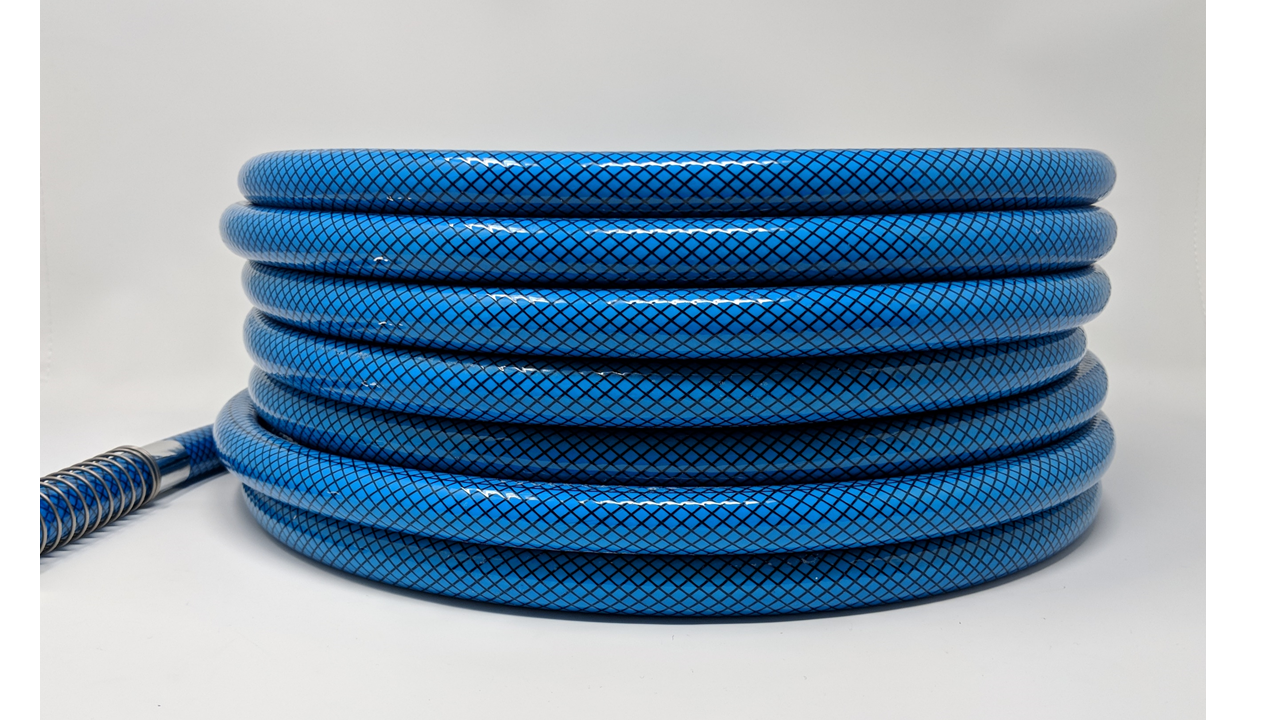}
\caption{ Composite water garden hose with reinforcing mesh.}
\label{fig_2}
\end{figure}

A popular current design consists of a polymer lining inside an elastomer
tube, containing helically wound and nearly inextensible filaments with
opposite helicity. A typical garden hose \cite{garden hose} with such
structure is shown in Fig.\ \ref{fig_2}.

An important element in such a hose is the reinforcing helical mesh which
plays a key role in the unusual behavior considered here. In all the
subsequent discussions, it will be assumed that the hose maintains its
circular cross-section everywhere, unless explicitly noted otherwise.

\section{The helical crossed fiber array constraint}

\begin{figure}[th]
\centering
\includegraphics[width=.8\linewidth]{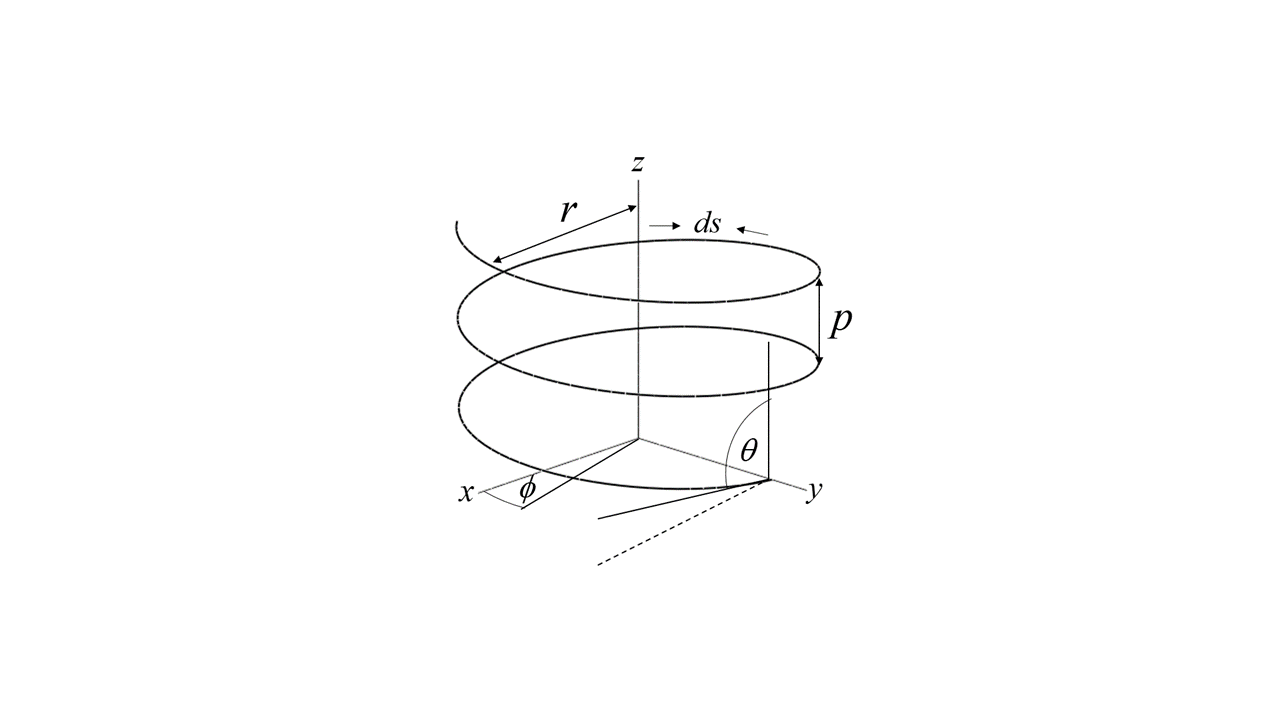}
\caption{Figure illustrating helix with identifying symbolics.}
\label{fig_3}
\end{figure}

Consider a helix, winding about the $z$-axis, as shown in Fig.\ \ref{fig_3}.
The position $\mathbf{r}$ of a point on the helix, in cylindrical
coordinates, is given by%
\begin{equation}
\mathbf{r}=(r\cos \phi ,r\sin \phi ,a\phi ),
\end{equation}%
where $r$ is the helix radius, and $a$ is related to the pitch of the helix,
$p$, via
\begin{equation}
p=2\pi a.
\end{equation}%
The length of the helix $s_{0}$ for one period can be obtained by
integrating the elemental arc length $ds$%
\begin{equation}
ds=\sqrt{r\sin ^{2}\phi d\phi ^{2}+r\cos ^{2}\phi d\phi ^{2}+a^{2}d\phi ^{2}}%
,
\end{equation}%
and the length of the helix for one period is
\begin{equation}
s_{0}=\sqrt{(2\pi r)^{2}+p^{2}}.  \label{a}
\end{equation}%
We note that we can write
\begin{equation}
\frac{p}{s_{0}}=\cos \theta  \label{11}
\end{equation}%
and%
\begin{equation}
\frac{2\pi r}{s_{0}}=\sin \theta ,  \label{12}
\end{equation}%
which defines $\theta $, the helix angle, the angle that the helix makes
with its axis.

We are interested in deformations of the hose allowed by the constraints due
to the reinforcing helical structure. The latter consists of pairs of fibers
forming helices of opposite handedness but equal pitch, as can be seen in
Fig. 2. Typically, a number of such fiber pairs are present, forming a
helical fiber array. This architecture resists twisting along the
longitudinal center line, since such a twist would reduce the pitch for one
of the helices, but increase it for the other. If there is no such twist and
the fibers are inextensible, then $s_{0}$ is a constant; it is the distance
along the helix for one period (where $\phi $ has changed by $2\pi $),
regardless of the pitch $p$. Since the total number of turns $N$ traced out
by either helix is a constant, $N$ is a constant of the hose. The length $L$
of the of hose is
\begin{equation}
L=Np,  \label{np}
\end{equation}%
and we then have that length of a fiber $L_{f}$ is
\begin{equation}
L_{f}^{2}=(N2\pi r)^{2}+L^{2},
\end{equation}%
which links the radius and the length of the hose. This is the principle
underlying the Chinese finger traps \cite{finger trap}; elongation causes
narrowing, and vice versa.

The volume of the cavity enclosed by the hose is
\begin{equation}
V=Np\pi r^{2},
\end{equation}%
where we have assumed, for simplicity, that the wall thickness is
negligible. Eliminating $r$ using Eq.\ \eqref{a}, we have%
\begin{equation}
V=\frac{N}{4\pi }s_{0}^{3}\left( \frac{p}{s_{0}}-\left( \frac{p}{s_{0}}%
\right) ^{3}\right) .  \label{V}
\end{equation}%
We see that changing the volume corresponds to changing the pitch and the
length of the hose as well as its radius.

We note that using Eqs.\ \eqref{11}, \eqref{np} and \eqref{V} we obtain
\begin{equation}
\frac{dV}{dL}=\frac{1}{4\pi }s_{0}^{2}(1-3\cos ^{2}\theta )  \label{dV}
\end{equation}%
and
\begin{equation}
\frac{dV}{dr}=\frac{N}{2}s_0^2\text{tan}\theta(3\text{cos}^2\theta-1 ).
\end{equation}
If
\begin{equation}
\cos \theta <\frac{1}{\sqrt{3}},
\end{equation}%
or if $\theta >54.74^{\circ}$, then increasing the volume lengthens and
narrows the hose, while if%
\begin{equation}
\cos \theta >\frac{1}{\sqrt{3}},
\end{equation}%
then increasing the volume shortens and widens the hose. In this way, the
helix angle determines whether the tube lengthens or shortens when its
volume is increased. At the critical helix angle $\theta _{c}=54.74^{\circ}$%
, and%
\begin{equation}
\cos \theta _{c}=\frac{1}{\sqrt{3}},
\end{equation}%
$dV/dL=0$, and the volume of the tube is maximum.

The measured helix angle of our garden hose is $\theta =36.9^{\circ}$,
placing it well into the regime where reducing the volume when the water
pressure is released results in elongation. The reason for the particular
choice of helix angle is not clear to us. (The manufacturer, Camco
Manufacturing, did not respond to our queries.)

\section{Elastic constants of the hose}

The elastic modulus of the hose is of paramount importance in determining
stability against buckling. We therefore consider the elastic energy cost of
shape deformations associated with changing the helix angle. The structure
of the hose material with the reinforcing helical filament array is clearly
anisotropic, and hence a full description of the elastic properties would
require at least 5 elastic constants \cite{5}. Here we provide the simplest
description consistent with the salient aspects of the elastic response.

We show that there are two distinct elastic constants in the system, and we
derive expressions enabling their experimental determination.

The elastic energy density of the hose associated with the deformation $%
\Delta \theta $ can be estimated in terms of the strains as \cite{elastic}%
\begin{equation}
\mathcal{E=}G(\varepsilon _{l}^{2}+\varepsilon _{a}^{2}+(\varepsilon
_{l}+\varepsilon _{a})^{2}),  \label{enr}
\end{equation}%
where $G$ is the shear modulus of the material in which the helical array is
embedded. We note that this calculation is based on constant volume of the
hose wall; the internal pressure is not considered.

The longitudinal strain, from Eq.\ \eqref{11} is%
\begin{equation}
\varepsilon _{l}=\frac{\Delta p}{p}=-\tan \theta \Delta \theta ,
\end{equation}%
and the azimuthal strain along the circumference of the hose, from Eq.\ %
\eqref{12} is
\begin{equation}
\varepsilon _{a}=\frac{\Delta r}{r}=\cot \theta \Delta \theta .
\end{equation}

The elastic energy density associated with deformations corresponding to
changing the helix angle, from Eq.\ \eqref{enr} is
\begin{equation}
\mathcal{E}=2G\left(\frac{1}{\sin ^{2}\theta \cos ^{2}\theta }%
-3\right)(\Delta \theta )^{2}.  \label{en}
\end{equation}

Writing the energy density in terms of the strains $\varepsilon _{l}$, and $%
\varepsilon _{a}$, we get
\begin{equation}
\mathcal{E}=2G\cot ^{2}\theta \left( \frac{1}{\sin ^{2}\theta \cos
^{2}\theta }-3\right) \varepsilon _{l}^{2} = \frac{1}{2}E_l\varepsilon_l^2,  \label{24}
\end{equation}%
and
\begin{equation}
\mathcal{E}=2G\tan ^{2}\theta \left( \frac{1}{\sin ^{2}\theta \cos
^{2}\theta }-3\right) \varepsilon _{a}^{2}= \frac{1}{2}E_a\varepsilon_a^2.  \label{25}
\end{equation}%
We can then regard the effective elastic modulus associated with elongation
of the hose from Eq.\ \eqref{24} as

\begin{equation}
E_{l}=4G\cot ^{2}\theta \left( \frac{1}{\sin ^{2}\theta \cos ^{2}\theta }%
-3\right)  \label{24e}
\end{equation}%
and the effective modulus associated with a corresponding change in the
radius from Eq.\ \eqref{25} as%
\begin{equation}
E_{a}=4G\tan ^{2}\theta \left( \frac{1}{\sin ^{2}\theta \cos ^{2}\theta }%
-3\right) .  \label{25e}
\end{equation}

We note that
\begin{equation}
E_{l}\tan ^{4}\theta =E_{a}.
\end{equation}%
Thus the elastic moduli are independent of wall thickness and radius, and $%
E_{l}$ and $E_{a}$ are both functions of the helix angle. The helical fiber
array acts as a transformer of the shear modulus $G$, giving rise to the two
effective elastic moduli. At $\theta =\theta _{c}$, $E_{l}=3G$, while $%
E_{a}=12G$. In the limiting case that $\theta =0$, $E_{l}=\infty $ and $%
E_{a}=4G$, while in the limit that $\theta =\pi /2$, $E_{l}=4G$ and $%
E_{a}=\infty $. It is important to stress, however, that thin flexible
fibers can support large stresses in tension, but not in compression.
Although an individual fiber would buckle readily in compression, this is
not the case when the fiber embedded in an elastic host. For simplicity we
neglect this contribution here, and assume for now that in compression the
material behaves as if no fiber was present. Hence the expressions for the
elastic constants in Eqs.\ \eqref{24e} and \eqref{25e} hold when the fibers
are in tension. In compression, the behavior is essentially that of a rubber
hose, and from Eq.\ \eqref{enr} we obtain
\begin{equation}
E_{l}=E_{a}=3G=E.  \label{30}
\end{equation}%
If $\theta <\theta _{c}$, then elongating the hose reduces its volume, while
if $\theta >\theta _{c}$ then elongating the hose increases the volume. \ If
the garden hose with constant volume and helix angle $\theta $ was under
longitudinal extensional stress, the helical fibers would be in tension so
long as $\theta <\theta _{c}$ and in compression otherwise, and if it was
under longitudinal compression, the fibers would be in tension so long as $%
\theta >\theta _{c}$, and in compression otherwise.

It is interesting to inquire about the stresses associated with the strains
in Eq.\ \eqref{enr}. Here, taking derivatives of the energy density with
respect to the strains to obtain the stresses is complicated by the fact
that the strains not independent. We therefore take the derivatives of the
energy density in Eqs.\ \eqref{24} and \eqref{25} and obtain%
\begin{equation}
\sigma _{l}=E_{l}\varepsilon _{l},
\end{equation}%
and
\begin{equation}
\sigma _{a}=E_{a}\varepsilon _{a}.
\end{equation}%
Taking the ratio, we find that the ratio of the stresses%
\begin{equation}
\frac{\sigma _{l}}{\sigma _{a}}=\frac{E_{l}}{E_{a}}\frac{\varepsilon _{l}}{%
\varepsilon _{a}}=\cot ^{2}\theta ,
\end{equation}%
which is $1/2$ at the critical angle. We remark here that $\sigma _{l}\,$is
the longitudinal stress required to produce the strain $\varepsilon _{l}$,
and similarly for $\sigma _{a}$ and $\varepsilon _{a}$, not stresses arising
from pressurizing the tube.

When the hose is pressurized, tension appears in the fibers. If the helix
angle is not critical, the ratio of the axial force due to the tension in the
fibers to the azimuthal force (easily calculated from the fiber density and
orientation) differs from $1/2$. This means that either the longitudinal or
the azimuthal force originating from pressure is unbalanced by tension in
the fibers, which leads to deformation of the hose, changing the helix angle
towards the critical value. This deformation is opposed by elastic stress in
the walls, but in the limit of high pressure, the helix angle reaches the
critical value. This is the principle of the McKibben actuator \cite%
{McKibben} where compressed air in a helical crossed fiber array is used as
artificial muscle in robotics and other applications.

The longitudinal elastic modulus of the garden hose may be determined by
simply stretching the empty hose without regard to its volume. Since the
stretching force $F$ is proportional to $\Delta L$,%
\begin{equation}
E_{l}=\frac{F}{2\pi rw}\frac{L}{\Delta L}.  \label{hel}
\end{equation}

Alternately, the longitudinal elastic modulus may be determined from the
volume change of the hose due to an applied pressure $P$. The hose volume $%
V=\pi rL$ can be written as%
\begin{equation}
V=\frac{N}{4\pi }s_{0}^{3}\cos \theta \sin ^{2}\theta ,
\end{equation}%
and the relative change in volume is

\begin{equation}
\frac{\Delta V}{V}=\frac{2-3\sin ^{2}\theta }{\cos \theta \sin \theta }%
\Delta \theta .  \label{rel}
\end{equation}

The relation between the relative volume change and the relative length
change is
\begin{equation}
\frac{\Delta V}{V}=\frac{1-3\cos ^{2}\theta }{1-\cos ^{2}\theta }\varepsilon
_{l}.  \label{long}
\end{equation}%
The total elastic energy of the hose is%
\begin{equation}
\mathcal{E}_{tot}=V_{mat}\mathcal{E=}2Ns_{0}^{2}wG\cos \theta \sin \theta
\left(\frac{1}{\sin ^{2}\theta \cos ^{2}\theta }-3\right)(\Delta \theta
)^{2}.
\end{equation}%
The pressure $P=$ $2\mathcal{E}_{tot}/\Delta V$, is then%
\begin{equation}
P=8G\frac{w}{r}\frac{(1-3\cos ^{2}\theta \sin ^{2}\theta )}{(2-3\sin
^{2}\theta )^{2}}\frac{\Delta V}{V},  \label{40}
\end{equation}%
which allows determination of the elastic constants.

Explicitly, from Eqs.\ \eqref{24e}, \eqref{25e} and \eqref{40} in terms of
the pressure and the volume change, we have%
\begin{equation}
E_{l}=\frac{1}{2}\frac{r}{w}\frac{P}{\Delta V/V}\left(3-\frac{2}{\sin
^{2}\theta }\right)^{2}  \label{wp}
\end{equation}%
and%
\begin{equation}
E_{a}=\frac{1}{2}\frac{r}{w}\frac{P}{\Delta V/V}\left(3-\frac{1}{\cos
^{2}\theta }\right)^{2}.
\end{equation}

\section{Hydroskeletons in nature}

Hydroskeletons consist of fluid-filled tubes, typically reinforced by
helical filaments of opposite handedness, with a variety of helix angles.
They are abundant in nature, ranging from sunflowers' stems to tubefeet of
starfishes, and shark skins \cite{Vogel, Kier}. Their utility lies in their
ability to become more or less rigid without rigid components.

If a helical fiber array-reinforced tube, with helix angle different from
the critical value, is filled with water at zero excess pressure, it will be
in stress-free equilibrium. If the ends are sealed, and the angle is less
than critical, the tube cannot be elongated, but a compressive force on the
ends can shorten the tube. Since this would increase the volume, the tube
cross-section cannot remain circular, and the tube will flatten. If the
angle is greater than critical, the tube cannot be shortened, but a tensile
force at the ends can elongate it. Again, this increases the volume, and the
tube must flatten. In either case, when the helix angle is not critical,
even though the tube with circular cross-section is completely filled with
an incompressible fluid, it will not be fully rigid, since it can become
flattened, and bend more easily. Flatworms, for example, make use of this
controllable rigidity in their locomotion \cite{Vogel}. However, when the
angle is critical, the volume is a maximum; neither elongation nor
shortening of the tube increases the volume, so the filled tube with closed
ends is nearly perfectly rigid.

Until fairly recently, it was believed that all hydroskeletons in nature
consisted of structures with helical fiber arrays \cite{Vogel}. However, in
1997, D.A. Kelly demonstrated that this is not the case. She showed that
reinforcing muscle fiber arrays in hydroskeletons in the penises of
armadillos were axial and orthogonal \cite{kelly1}; that is, either parallel
to the tube axis, or perpendicular to it. (Strictly speaking, there are \textbf{%
two} helical arrays, with the limiting cases $\theta =0$ and $\theta =\pi /2$%
. The corresponding elastic constants $E_{l}$ and $E_{a}$ are then
infinite.) Other realizations of such fiber arrays have since been
identified \cite{Kier}. We return to the questions of the utility of this
rare architecture in Section 10.

Before concluding this section on helical crossed-fiber arrays, we return to
blood vessels: arteries. Arteries are by no means thin-walled; typically $%
r/w=2$. A crude estimate of the longitudinal strain due to the systolic and
diastolic pressure difference is $\varepsilon _{l}\simeq 1\%$. However, no
such strain is observed. This is likely due to the structure of the artery
walls: two of the three layers constituting the wall contain collagen
helical filament arrays, likely near the critical angle \cite{Holzapfel}.

Finally, we note that osteons, the building blocks of dense bone, have
cylindrical lamellar structures with embedded crossed helical collagen
fibers \cite{MM}. It appears that helical fiber arrays abound in nature;
they are practically everywhere. This makes the existence of the recently
discovered axial-orthogonal architecture \cite{kelly1} even more remarkable.

\section{Buckling and wavelength selection}

Ignoring the effects of gravity, if the hose is full of water, but under no
excess pressure, it is undeformed, with zero stress and strain. If the hose
is then pressurized, the volume of the hose increases, in our case, since $%
\theta <\theta _{c}$, by getting wider and shorter. The hose has to undergo
an elastic deformation, subject to the inextensibility constraint of the
helical fibers. Elastic energy is stored in the deformed strained hose wall.
If the inlet valve is next closed and the spray nozzle is opened (or the
hose simply leaks at a joint), water leaves the hose, the pressure drops,
and the hose elongates. The observed buckling is a direct consequence of the
elongation of the hose.

To understand the proposed wavelength selection mechanism, we first imagine
a straight section of the hose of length $2l$ and consider its elongation
assuming that the ends are free to move. Due to symmetry, the center of mass
will not move during the elongation; instead the two ends will move away
from the center. Since the hose is full of water, it has considerable
weight, and there will be kinetic friction between the hose and the
grass/earth supporting its weight. Each element of length of the hose must
have a net force acting on it to overcome the kinetic friction. The
longitudinal stress in the hose due to friction must therefore be a linear
function of position: maximum in the middle, decreasing linearly and
vanishing at the ends. If $\mu _{k}$ is the kinetic friction coefficient,
the friction force per unit length acting on the hose is
\begin{equation}
f=\mu _{k}\rho g,
\end{equation}%
where $\rho $ is the linear mass density and $g$ is the acceleration of
gravity. The longitudinal stress in the walls of the tube at the midpoint
must therefore be
\begin{equation}
\sigma _{l}=\mu _{k}\rho gl/(2\pi rw).
\end{equation}%
Clearly, stress in the wall increases with distance from the end, and there
must be a point where the tube becomes unstable against buckling.

If only one-half of the tube is considered, with stress linearly increasing
with distance from the free end, its stress is equivalent to that in a
column loaded by its own weight. The stability of such a clamped-free column
has been examined by Euler \cite{Euler1, Euler2, Euler3}. We remark here
that we have not carried out full stability analysis of the elongating
garden hose, instead we rely on the analogy with self-buckling of a column.
The analogy is imperfect, since the the garden hose is anisotropic due to
the fiber structure, however, we believe that the mechanism for the buckling
is correctly identified. We also note that the irregularities in grass
density, height and ground elevation will impact on the local friction
force, and introduce perturbations with consequences that are not well
understood.

Remarkably, Euler erred in his first calculation \footnote{%
For history see Wikipedia entry on self-buckling.}, but it was subsequently shown by Dinnik
\cite{Dinnik} that the column will buckle at the critical length $l_{c}$%
\begin{equation}
l_{c}^{2}=\alpha \frac{EI}{F_{c}},
\end{equation}%
where $\alpha =7.8373$, $F_{c}$ is the total load, $E$ is Young's modulus
and $I$ is the moment of area of the cross-section. In our example,
\begin{equation}
F_{c}=\mu _{k}\rho gl_{c}
\end{equation}%
and we obtain
\begin{equation}
l_{c}^{3}=\alpha \frac{EI}{\mu _{k}\rho g}.
\end{equation}%
We therefore expect the wavelength of the deformation $\lambda =2l_{c}$, or%
\begin{equation}
\lambda =2\left(\alpha \frac{EI}{\mu _{k}\rho g}\right)^{1/3} = 2\left(\alpha \frac{EI}{f}\right)^{1/3}. \label{lamb}
\end{equation}%
This is our main result. Even though it is obtained via analogy and the
prefactor may not be exact, we expect the scaling aspects to hold.

\section{Amplitude of the sinusoidal undulation}

If the buckling mechanism is as proposed above, then the amplitude of the
sinusoidal deformation is simply related to the elongation of the hose; the
length of the deformed hose in one period must equal the wavelength times
the longitudinal stretch $1+\varepsilon _{l}$.

If the shape of the curve traced out by the hose is sinusoidal, that is, of
the form%
\begin{equation}
y=A\cos (2\pi \frac{x}{\lambda }),
\end{equation}%
then
\begin{equation}
\lambda (1+\varepsilon _{l})=\int_{0}^{\lambda }\sqrt{y^{\prime 2}+1}dx.
\end{equation}%
where $y^{\prime }=d y/d x$.

For small $y^{\prime }$, we have
\begin{eqnarray}
\lambda (1+\varepsilon _{l}) &\simeq &\int_{0}^{\lambda }1+\frac{1}{2}%
A^{2}\left((2\pi \frac{1}{\lambda })^{2}\sin ^{2}(2\pi \frac{x}{\lambda }%
)\right)dx  \notag \\
&=&\lambda \left(1+\frac{1}{4}A^{2}\left(2\pi \frac{1}{\lambda }%
\right)^{2}\right)
\end{eqnarray}%
and the amplitude is given, approximately, by%
\begin{equation}
A\simeq \frac{1}{\pi }\lambda \sqrt{\varepsilon _{l}}.  \label{amp}
\end{equation}

\section{Experimental results}

The physical dimensions of the hose were determined by dissecting the hose
and measuring the diameter and wall thickness. The outside diameter was
found to be $20$ mm and the inside diameter $16$ mm.

The moment of area of a thin tube is%
\begin{equation}
I=\pi r^3w,
\end{equation}%
where $w$ is the wall thickness. In our case,
\begin{equation}
I=3.27\times 10^{-9}\text{ m}^{4}.
\end{equation}

Key quantities determining the wavelength of the instability are the elastic
modulus and the kinetic friction.

The friction force per length $f$ was measured a number of times by dragging
a length of hose and measuring the required force with a spring balance. The
required force can vary considerably due to different conditions of
humidity, temperature and grass height, wetness and stiffness. We found%
\begin{equation}
f=27\pm 4\text{ N/m}.
\end{equation}

Another key quantity is the relevant elastic constant. In Euler's original
self-buckling calculation \cite{Euler3, Dinnik}, an isotropic material was
considered. Our hose is anisotropic, with two different effective elastic
constants. We assumed that the relevant elastic constant in our case is $%
E_{l} $ which relates the total elastic energy to changes in length, since
both bulk compression and bend are associated primarily with longitudinal
strains.

The first and most direct method to determine the longitudinal elastic
constant $E_{l}$ is to stretch the empty hose by applying a force manually,
and to measure the displacement $\Delta L$ of the end of the hose with a
meter stick, and the force with a spring balance. We found that the
displacement $\Delta L$ was proportional to the force $F$,
\begin{equation}
\frac{F}{\Delta L}=73.1\text{ N/m},
\end{equation}%
and for our $L=30.1$ m hose, Eq.\ \eqref{hel} gives%
\begin{equation}
E_{l}=22.0\text{ MPa}.
\end{equation}%
We remark here that the force required to stretch the hose by fixed length
is time-dependent: there is a relaxation time on the scale of seconds. We
took force readings $\sim 10$ s after the displacement when the force
reading was nearly constant.

The second and less direct method of determining the longitudinal elastic
constant consisted of filling the hose with pressurized water and measuring
the total volume $V$ of the water in the hose. Subtracting from this volume
the volume of the empty hose gives the extra volume $\Delta V$ of the hose
due to pressure $P$. We found that from measurements on a $30.1$ m hose that
\begin{equation}
\frac{\Delta V}{V}=0.20\pm 0.06.
\end{equation}%
There is considerable uncertainty in these measurements due to errors
associated in part with the dynamic response of the hose, and in part with
accurately measuring the volume of water in the hose. The water pressure at
supply was $P=426$ kPa, and from Eq.\ \eqref{wp} we obtain

\begin{equation}
E_{l}=27.8\text{ MPa,}
\end{equation}%
in fair agreement with the result from the stretching method. We assumed that
a reasonable estimate of the effective longitudinal elastic modulus is the
average of the two measurements, $\langle E_{l}\rangle =24.9$ MPa.

Now, when a cylindrical column or hose begins to buckle, the neutral surface
is along the symmetry axis of the hose, and one half of the hose is in
tension, while the other is in compression. Since the helical fibers do not
significantly contribute to the stress in compression, the elastic constant
of the hose in the compressed half is Young's modulus. We did not carry out
elastic constant measurements under compression; however, from Eq.\ %
\eqref{30} we obtain for Young's modulus
\begin{equation}
E=3G=13.8\text{ MPa.}
\end{equation}%
We assume that the effective elastic constant for buckling is the average of
these two; that is,
\begin{eqnarray}
E_{eff} &=&\frac{1}{2}(E_{l}+E)  \notag \\
&=&19.4\text{ MPa.}
\end{eqnarray}

We note that longitudinal strain under pressure can be obtained from Eq.\ %
\eqref{long} to give%
\begin{equation}
\varepsilon _{l}=0.078.
\end{equation}

We can now calculate the wavelength of the modulation. We recall our main
result, Eq.\ \eqref{lamb}, duplicated here for convenience,
\begin{equation}
\lambda =2\left(\alpha \frac{E_{eff}I}{f}\right)^{1/3},
\end{equation}
which gives the theoretical prediction
\begin{equation}
\lambda _{th}=0.53\text{ m}.
\end{equation}

\begin{figure}[th]
\centering
\includegraphics[width=.5\linewidth]{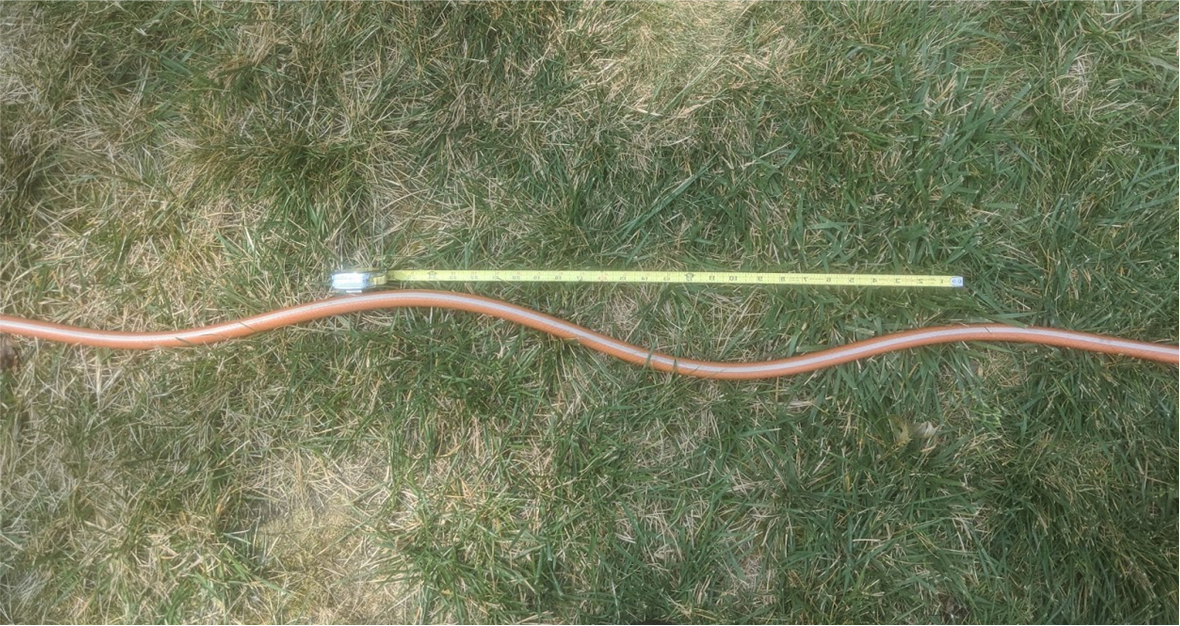}
\caption{A typical example of an undulation with $\protect\lambda =66$ cm. }
\label{fig_4}
\end{figure}

Our measured values have considerable scatter, likely due again to
variations in humidity, temperature and grass conditions. The average
experimental value is%
\begin{equation}
\lambda _{\exp }=0.64\pm 0.16\text{ m}.
\end{equation}%
A typical example is shown in Fig.\ \ref{fig_4}.

The amplitudes of the sinusoidal undulations, which were clearly affected by
surface topography, were not studied carefully. The theoretical estimate
from Eq.\ \eqref{amp} is

\begin{equation}
A=0.05\text{ m},
\end{equation}%
and from our observations (see Fig. 3), we estimate%
\begin{equation}
A_{\exp }=0.05\pm 0.04\text{ m}.
\end{equation}%
Model predictions are therefore in good agreement with experiment.

\section{A unique feature of the axial orthogonal architecture}

Before concluding, we consider again the more recently discovered axial
orthogonal fiber architecture \cite{kelly1} in the penises of mammals. In
the natural world where helical fiber arrays abound, the existence of such a
singular structure suggests a special and valuable property arising from it.
While numerous studies \cite{Koehl, KellyDA} claim that the axial orthogonal
arrangement has greater resistance to bending than helical arrays, we have
been unable to find measurements verifying this. There also appears to be no
clear explanation of how axial orthogonal fibers create more resistance to
bending than helical ones.

It is clear that ideal hydroskeletons, with inextensible fibers arranged
either as helical arrays at the critical angle or axial orthogonal
arrangements at maximum volume, cannot be deformed. Since deformations exist
in nature, either the fibers must be extensible, or the constant maximum
volume constraint must be relaxed. As in the stability analysis of our
garden hose, we consider here inextensible (but not incontractable) fibers,
and relax the constant volume constraint.

One biologically important characteristic of mammalian penises is stability
against buckling \cite{Udelson, Wassersug}.

We first consider the buckling of a straight piece of garden hose of length $%
l$ reinforced with the helical crossed fiber array having the critical helix
angle $\theta _{c}$ under a load $F_{h}$. In buckling, at the critical
force, the hose starts to bend. The neutral surface, whose area is unchanged
as the hose starts to bend is along the axis of the hose. On one side of
this surface, the hose is in tension; in the other, in compression. As in a
column, the critical force for buckling is
\begin{equation}
F_{hc}=\pi ^{2}\frac{EI}{l^{2}},
\end{equation}%
where $E=3G$ is Young's modulus of the hose wall. (At the critical helix
angle, the longitudinal elastic constant is $E_{l}=3G=E$ both in tension and
compression.)

To understand buckling with the axial orthogonal structure, it is useful to
consider the parallels with the bending of a beam with rectangular cross
section. If the beam is uniform, one surface will be stretched, the opposite
surface will be compressed. The neutral surface, with an area that is
unchanged as the beam is bent, is at the midplane between these two surfaces.
If one now considers the same beam modified so that the previously stretched
surface is now made inextensible (say by attaching inextensible axial filaments
to the surface) and the beam is similarly bent, the inextensible surface will
become the neutral surface. This translation of the neutral surface from the
midplane to the beam surface increases the moment of area\footnote{%
This is similar to the change of the moment of inertia when the axis of
rotation is translated.} according to the parallel axis (Huygens-Steiner)
theorem \cite{steiner}, and the bending moment needed to create the same
bend will increase by a factor of $4$ for a beam with rectangular
cross-section.

Returning to the hose with the circular cross-section, the critical force
for buckling with the axial orthogonal fiber array is%
\begin{equation}
F_{ac}=3\pi ^{2}\frac{EI}{l^{2}}.
\end{equation}%
Due to the inextensibility of the axial fibers, the neutral plane has moved
from the center to the outside surface of the wall, the moment of area has
increased by a factor of $3$ for a thin walled tube (by a factor of $5$ for
a solid cylinder), and the walls are in compression with Young's modulus $E$%
. The axial orthogonal fiber arrangement thus provides a significantly
increased stability against both bending and buckling due to the increased
the moment of area. The effect is caused by the axial inextensibility of the
outside of the wall due to the axial orthogonal arrangement; a property
unique to this structure. We suppose this effect to persist, in essence,
even when the hydroskeleton is full. We suggest therefore that the key
property furnished by the axial orthogonal array is the approximately
three-fold increase in stability against buckling; and look forward to
experiments for quantitative assessment of this prediction. Later stage
structural deformations beyond threshold also merit study.

\section{Conclusions}

We have studied the spontaneous buckling instability exhibited by certain
garden hoses. We have found that the instability occurs as a result of hose
elongation when water is released, and the internal pressure is reduced. The
elongation is associated with the constraint on shape change associated with
a helical array of fibers, with helix angles less than $\theta
_{c}=54.7^{\circ}$. During elongation, kinetic friction between the hose and
the supporting surface results in stress buildup in the hose, leading to
buckling. The instability is similar to the self-buckling of columns,
enabling prediction of the selected wavelength. This mechanism enables
simple classroom demonstrations of self-buckling using friction force as the
load.

A simple model for the elastic response of the hose enabled determination of
the relevant elastic modulus of the hose. This, together with the measured
friction force, allowed comparison of the predicted and observed wavelengths
and undulation amplitudes. We have found reasonable agreement between theory
and experiment. In addition to predicting the instability and its details,
our simple model provides useful insights into the effects of helical fiber
arrays on the material's response. The constraint of the array couples
longitudinal and azimuthal strains, and the helix angle changes the
effective longitudinal and azimuthal elastic moduli of the otherwise
isotropic tube. The ability to tune elastic moduli in this way promises to
be useful in the design of active shape-changing materials, such as
photomechanical elastomers for a variety of applications. For example, a
rubber cylinder encased in a helical crossed fiber array of photomechanical
filaments would transform filament contraction to robust extension. Finally,
a simple mechanism has been proposed to explain the high buckling resistance
afforded by the axial orthogonal fiber arrangement in the penises of mammals.

\section{Acknowledgements}

We are grateful for illuminating discussions with Diane Kelly ( UMass Amherst) and Mimi A.R. Koehl (UC Berkeley).
This work was supported by the Office of Naval Research [ONRN00014-18-1-2624].

\end{document}